\documentclass[aps,prd,superscriptaddress]{revtex4}
\usepackage{epsfig,epsf}
\usepackage{amsmath}
\usepackage{amsthm}
\usepackage{amsfonts}
\usepackage{amssymb}
\usepackage{dsfont}
\usepackage{multirow}
\usepackage{appendix}
\usepackage{slashed}
\usepackage[active]{srcltx}
\usepackage{psfrag}

\begin{document}

\title{Molecular states $J/\psi B_{c}^{+}$ and $\eta_{c}B_{c}^{\ast +} $}
\date{\today}
\author{S.~S.~Agaev}
\affiliation{Institute for Physical Problems, Baku State University, Az--1148 Baku,
Azerbaijan}
\author{K.~Azizi}
\thanks{Corresponding Author}
\affiliation{Department of Physics, University of Tehran, North Karegar Avenue, Tehran
14395-547, Iran}
\affiliation{Department of Physics, Dogus University, Dudullu-\"{U}mraniye, 34775
Istanbul, T\"{u}rkiye}
\author{H.~Sundu}
\affiliation{Department of Physics Engineering, Istanbul Medeniyet University, 34700
Istanbul, T\"{u}rkiye}

\begin{abstract}
Hadronic molecules $\mathfrak{M}=J/\psi B_{c}^{+}$ and $\widetilde{\mathfrak{%
\ M}}=\eta _{c}B_{c}^{\ast +}$ are investigated in the framework of QCD sum
rule method. These particles with spin-parities $J^{\mathrm{P}}=1^+$ have
the quark contents $cc \overline{c}\overline{b}$. We compute their masses
and current couplings and find that they are numerically very close to each
other coinciding within accuracy of the sum rule method. Therefore, we
concentrate on the molecule $J/\psi B_{c}^{+}$ and explore features of this
state in a detailed form. Our prediction $m=(9740 \pm 70)~\mathrm{MeV}$ for
its mass means that $\mathfrak{M}$ easily decays to pairs of ordinary mesons
through strong interactions. There are two mechanisms responsible for
transformations of $\mathfrak{M}$ to conventional mesons. The fall-apart
mechanism generates the dominant decay channels $\mathfrak{M} \to J/\psi
B_{c}^{+}$ and $\mathfrak{M} \to \eta _{c}B_{c}^{\ast +}$. Annihilation of $%
\overline{c}c$ quarks triggers subdominant processes with various
final-state $B$ and $D$ mesons: Six of such channels are considered in this
work. The partial widths all of decays are computed using the three-point
sum rule approach. The width $\Gamma[ \mathfrak{M}]=(121 \pm 17)~ \mathrm{MeV%
}$ of the hadronic axial-vector molecule $\mathfrak{M}$, as well as its mass
provide valuable information for running and future experiments.
\end{abstract}

\maketitle


\section{Introduction}

\label{sec:Intro}

Hadronic molecules built of conventional meson pairs are interesting objects
and deserve detailed studies within framework of high energy physics. Such
molecules may be composed of two mesons and contain four quarks of the same
or different flavors. Because existence of such systems, as well as ones
with diquark-antidiquark organization, does not contradict to principles of
the Quantum Chromodynamics (QCD), their theoretical and experimental
analyses are among priorities of the hadron physics.

Suggestions to consider some of resonances observed in experiments as
bound/resonant states of conventional mesons were made fifty years ago \cite%
{Bander:1975fb,Voloshin:1976ap,DeRujula:1976zlg}. Thus, it was supposed to
interpret numerous vector states discovered in $e^{+}e^{-}$ annihilations as
structures composed of $D$ meson pairs. These hypothetical molecules contain
both the heavy and light quarks.

In the years since, theoretical studies led to advancing the idea of
hadronic molecules. Now information collected on parameters of numerous
hadronic molecules, on their binding, production and decay mechanisms forms
rather strong basis for running and future experiments. This theoretical
activity was accompanied by elaborating of new and adapting existing methods
and models for investigations of hadronic molecular states. Last
publications and references therein give one a better understanding of
achievements made in this growing area of the hadron physics \cite%
{Molina:2020hde,Xu:2020evn,Xin:2021wcr,Wang:2023bek,Agaev:2022duz,Agaev:2023eyk,Braaten:2023vgs,Wu:2023rrp, Liang:2023jxh,Wang:2025zss,Yalikun:2025ssz,Liu:2023gla,Liu:2024pio}%
.

Fully heavy hadronic molecules, i.e., molecules composed of heavy mesons and
containing only $b$ and $c$ quarks are relatively new states attracted
interests due to recent LHCb-CMS-ATLAS discoveries of $X$ structures \cite%
{LHCb:2020bwg,ATLAS:2023bft,CMS:2023owd}. These presumably exotic $cc%
\overline{c}\overline{c}$ mesons were investigated by employing different
approaches. The diquark-antidiquark and hadronic molecule pictures are
mostly utilized models to describe properties of these states. Relevant
problems were also addressed in our articles Refs.\ \cite%
{Agaev:2023ruu,Agaev:2023rpj}, in which we computed parameters of the
molecules $\eta _{c}\eta _{c}$, $\chi _{c0}\chi _{c0}$, and $\chi _{c1}\chi
_{c1}$ and interpreted some of them as candidates to $X$ structures.

The heavy charm-bottom molecules built of equal number of $b$ and $c$ quarks
were examined in different papers \cite%
{Liu:2023gla,Liu:2024pio,Wang:2023bek,Agaev:2025wdj,Agaev:2025fwm,Agaev:2025nkw}%
. The coupled-channel unitary approach was applied in Ref.\ \cite%
{Liu:2023gla} to study states $B_{c}^{(\ast )+}B_{c}^{(\ast )-}$. The
parameters of such molecules with $J^{\mathrm{PC}}=0^{++}$, $1^{++}$, and $%
2^{++}$ were calculated in our works \cite%
{Agaev:2025wdj,Agaev:2025fwm,Agaev:2025nkw}. We performed corresponding
analyses using the QCD sum rule (SR) method, which allows one to evaluate
parameters of ordinary and exotic hadrons \cite%
{Shifman:1978bx,Shifman:1978by,Albuquerque:2018jkn,Agaev:2020zad}. In this
approach we find the mass $m\pm \delta m$ of a particle with the central
value $m$ and ambiguities $\pm \delta m$ emerged during these computations.
In other words, the region $[m-\delta m$, $m+\delta m]$ can be used to fix
the mass of a particle of interest.

The fully heavy asymmetric molecules with contents $bb\overline{b}\overline{c%
}$ and $cc\overline{c}\overline{b}$ were explored in the literature \cite%
{Liu:2024pio,Agaev:2025did,Agaev:2025wyf} as well. In Refs.\ \cite%
{Agaev:2025did,Agaev:2025wyf} the scalar $\mathcal{M}_{\mathrm{b}}=\eta
_{b}B_{c}^{-}$ and $\mathcal{M}_{\mathrm{c}}=\eta _{c}B_{c}^{+}$, and
axial-vector $\mathcal{M}_{\mathrm{AV}}=\Upsilon B_{c}^{-}$ [$\widetilde{%
\mathcal{M}}_{\mathrm{AV}}=\eta _{b}B_{c}^{\ast +}$] structures were studied
using the SR method, where we computed the masses and decay widths of these
structures. The molecules with a prevalence of $b$-quarks are close to
corresponding two-meson thresholds provided one uses for analysis the
central value $m$. In this scenario, both the scalar $\mathcal{M}_{\mathrm{b}%
}$ and axial-vector $\mathcal{M}_{\mathrm{AV}}$ molecules are resonant
states and dissociate to constituent mesons. In the lower limit of the
masses, i.e., in a situation when $m-\delta m$, the molecules $\mathcal{M}_{%
\mathrm{b}}$ and $\mathcal{M}_{\mathrm{AV}}$ reside below two-meson
thresholds and form bound structures. This means that they cannot decay to
mesons which are their ingredients, or decay to particles which contain all
four initial quarks. Nevertheless, $\mathcal{M}_{\mathrm{b}}$ and $\mathcal{M%
}_{\mathrm{AV}}$ transform to ordinary mesons through strong processes which
occur due to annihilation of $b\overline{b}$ quarks into light $q\overline{q}
$ and $s\overline{s}$ pairs followed by creations of ordinary mesons \cite%
{Becchi:2020mjz,Becchi:2020uvq,Agaev:2023ara}. Thus, there are two
mechanisms for transformation of four-quark hadronic molecules to ordinary
particles. In our previous works \cite{Agaev:2025did,Agaev:2025wyf} we used
both of these mechanisms to evaluate total widths of the hadronic molecules $%
\eta _{b}B_{c}^{-}$ and $\Upsilon B_{c}^{-}$.

In the case of the scalar molecule $\mathcal{M}_{\mathrm{c}}=\eta
_{c}B_{c}^{+}$ this scenario changes qualitatively. It turns out that the
molecule $\mathcal{M}_{\mathrm{c}}$ dissociates to its components $\eta
_{c}+B_{c}^{+}$ even in the lower value of its mass. Hence, it does not form
a bound state in a sense emphasized above.

In present work, we study the axial-vector counterparts of $\mathcal{M}_{%
\mathrm{c}}$, i.e., consider the molecules $\mathfrak{M}=J/\psi B_{c}^{+}$
and $\widetilde{\mathfrak{M}}=\eta _{c}B_{c}^{\ast +}$ built of $cc\overline{%
c}\overline{b}$ quarks. It is evident that these quarks may be grouped into
two different colorless components-ordinary heavy mesons. Thus, $\mathfrak{M}
$ is a system of the vector charmonium $J/\psi $ and pseudoscalar $B_{c}^{+}$
meson, whereas $\widetilde{\mathfrak{M}}$ contains the pseudoscalar
charmonium $\eta _{c}$ and vector meson $B_{c}^{\ast +}$. We are going to
compute the masses and decay widths of $\mathfrak{M}$ and $\widetilde{%
\mathfrak{M}}$ in the framework of QCD SR method. We find that whilst $%
\mathfrak{M}$ and $\widetilde{\mathfrak{M}}$ have different internal
organizations and are different molecules, their masses and current
couplings are numerically very close each other. They have identical quark
contents and quantum numbers, as a result, their decay patterns do not
differ from each other as well. Because the SR method leads to predictions
with some uncertainties, we cannot clearly distinguish these structures.
Therefore, in this work we concentrate on properties of the molecule $%
\mathfrak{M}$.

To this end, apart from the mass and current coupling, we address also
various decay channels of the molecule $\mathfrak{M}$. It is
strong-interaction unstable state and decays to ordinary meson pairs through
two mechanisms discussed just above. The processes $\mathfrak{M\rightarrow }%
J/\psi B_{c}^{+}$ and $\eta _{c}B_{c}^{\ast +}$ are dominant decay modes of
the molecule $\mathfrak{M}$. The second mechanism gives rise to subdominant
channels $\mathfrak{M\rightarrow }B^{\ast +}D^{0}$, $B^{\ast 0}D^{+}$, $%
B^{+}D^{\ast 0}$,$B^{0}D^{\ast +}$, $B_{s}^{\ast 0}D_{s}^{+}$, and $%
B_{s}^{0}D_{s}^{\ast +}$ which belong to this category of processes.

To evaluate the partial widths all of aforementioned decays, we invoke
technical tools of the three-point SR method necessary to find strong
couplings at corresponding $\mathfrak{M}$-meson-meson vertices. When
exploring the annihilation channels we employ also a relation between heavy
quark vacuum expectation value $\langle \overline{c}c\rangle $ and gluon
condensate $\langle \alpha _{s}G^{2}/\pi \rangle $. But this manipulation
does lead to additional free parameters in analysis.

This article is organized in the following way: The masses and current
couplings of the hadronic molecules $J/\psi B_{c}^{+}$ and $\eta
_{c}B_{c}^{\ast +}$ are calculated in Sec.\ \ref{sec:Mass}. In Sec.\ \ref%
{sec:Widths1} we consider the decay channels $\mathfrak{M\rightarrow }J/\psi
B_{c}^{+}$ and $\eta _{c}B_{c}^{\ast +}$ and calculate their partial widths.
The subdominant modes of $\mathfrak{M}$ are explored in Secs. \ref%
{sec:Widths2} and \ref{sec:Widths3}. In section \ref{sec:Widths3} we present
our prediction for the full decay width of the molecule $J/\psi B_{c}^{+}$.
The last part, \ref{sec:Conc}, of this paper is reserved for discussion and
contains also our concluding remarks.


\section{Masses and current couplings of the molecules $J/\protect\psi %
B_{c}^{+}$ and $\protect\eta _{c}B_{c}^{\ast +}$}

\label{sec:Mass}

To find the spectroscopic parameters of the hadronic molecules $J/\psi
B_{c}^{+}$ and $\eta _{c}B_{c}^{\ast +}$, one should start form analysis of
the correlator
\begin{equation}
\Pi _{\mu \nu }(p)=i\int d^{4}xe^{ipx}\langle 0|\mathcal{T}\{I_{\mu
}(x)I_{\nu }^{\dag }(0)\}|0\rangle ,  \label{eq:CF1}
\end{equation}%
where $I_{\mu }(x)$ [$\widetilde{I}_{\mu }(x)$] is the interpolating current
for the molecule $\mathfrak{M}$ ($\widetilde{\mathfrak{M}})$, while $%
\mathcal{T}$ is the time-ordered product of currents. The current $I_{\mu
}(x)$ has the following form
\begin{equation}
I_{\mu }(x)=\overline{c}_{a}(x)\gamma _{\mu }c_{a}(x)\overline{b}%
_{b}(x)i\gamma _{5}c_{b}(x).
\end{equation}%
In the case of the structure $\widetilde{\mathfrak{M}}$, we employ the
current
\begin{equation}
\widetilde{I}_{\mu }(x)=\overline{c}_{a}(x)i\gamma _{5}c_{a}(x)\overline{b}%
_{b}(x)\gamma _{\mu }c_{b}(x),
\end{equation}%
and calculate the correlator $\widetilde{\Pi }_{\mu \nu }(p)$ obtained from
Eq.\ (\ref{eq:CF1}) after substitution $I_{\mu }(x)\rightarrow \widetilde{I}%
_{\mu }(x)$. In these currents $b(x)$ and $c(x)$ are the heavy quark fields
with color indices $a$ and $b$.

To find sum rules for the mass $m$ and current coupling $\Lambda $ of $%
\mathfrak{M}$ one has to calculate the correlation function $\Pi _{\mu \nu
}(p)$ using two different approaches. First, it should be expressed in terms
of the parameters $m$ and $\Lambda $, which establishes the phenomenological
component $\Pi _{\mu \nu }^{\mathrm{Phys}}(p)$ of the relevant SRs. The
correlation function $\Pi _{\mu \nu }^{\mathrm{Phys}}(p)$ is given by the
formula
\begin{equation}
\Pi _{\mu \nu }^{\mathrm{Phys}}(p)=\frac{\langle 0|I_{\mu }|\mathfrak{M}%
(p,\epsilon )\rangle \langle \mathfrak{M}(p,\epsilon )|I_{\nu }^{\dagger
}|0\rangle }{m^{2}-p^{2}}+\cdots ,  \label{eq:Phys1}
\end{equation}%
with $\epsilon _{\mu }$ being the polarization vector of the axial-vector
state $\mathfrak{M}$. We write only down the contribution of the
ground-state molecule, while the dots stand for contributions of the higher
resonances and continuum states.

To compute $\Pi _{\mu \nu }^{\mathrm{Phys}}(p)$ we make use of the matrix
element
\begin{equation}
\langle 0|I_{\mu }|\mathfrak{M}(p,\epsilon )\rangle =\Lambda \epsilon _{%
\mathcal{\mu }}(p),  \label{eq:ME1}
\end{equation}%
and get
\begin{equation}
\Pi _{\mu \nu }^{\mathrm{Phys}}(p)=\frac{\Lambda ^{2}}{m^{2}-p^{2}}\left(
g_{\mu \nu }-\frac{p_{\mu }p_{\nu }}{m^{2}}\right) +\cdots .
\label{eq:Phys2}
\end{equation}

The correlator $\Pi _{\mu \nu }(p)$ has to be found in terms of the heavy
quark propagators $S_{b(c)}(x)$ (see, Ref.\ \cite{Agaev:2020zad}) and
calculated by applying the techniques of the operator product expansion ($%
\mathrm{OPE}$). For $\Pi _{\mu \nu }^{\mathrm{OPE}}(p)$, we obtain
\begin{eqnarray}
\Pi _{\mu \nu }^{\mathrm{OPE}}(p)&=&i\int d^{4}xe^{ipx}\mathrm{Tr}\left\{ %
\left[ \gamma _{\mu }S_{c}^{ab^{\prime }}(x)\gamma _{5}S_{b}^{b^{\prime
}b}(-x)\gamma _{5}S_{c}^{ba^{\prime }}(x)\gamma _{\nu }S_{c}^{a^{\prime
}a}(-x)\right] \right.  \notag \\
&&\left. -\mathrm{Tr}\left[ \gamma _{\mu }S_{c}^{aa^{\prime }}(x)\gamma
_{\nu }S_{c}^{a^{\prime }a}(-x)\right] \mathrm{Tr}\left[ \gamma
_{5}S_{c}^{bb^{\prime }}(x)\gamma _{5}S_{b}^{b^{\prime }b}(-x)\right]
\right\} ,  \label{eq:QCD1}
\end{eqnarray}%
which is the QCD side of the sum rules.

The correlation functions $\Pi _{\mu \nu }^{\mathrm{Phys}}(p)$ and $\Pi
_{\mu \nu }^{\mathrm{OPE}}(p)$ are composed of two Lorentz structures. The
terms $\sim g_{\mu \nu }$ in these correlators are formed due to
contributions of the spin-1 particle, therefore are suitable for SR
analysis. We denote by $\Pi ^{\mathrm{Phys}}(p^{2})$ and $\Pi ^{\mathrm{OPE}%
}(p^{2})$ corresponding invariant amplitudes. In the case of $\Pi _{\mu \nu
}^{\mathrm{Phys}}(p)$ this amplitude is equal to $\Pi ^{\mathrm{Phys}%
}(p^{2})=\Lambda ^{2}/(m^{2}-p^{2})$.

Having equated $\Pi ^{\mathrm{Phys}}(p^{2})$ and $\Pi ^{\mathrm{OPE}}(p^{2})$%
, performed required manipulations (see, for instance, Ref. \cite%
{Agaev:2024uza}), we derive SRs for the mass $m$ and current coupling $%
\Lambda $ of the molecule $\mathfrak{M}$
\begin{equation}
m^{2}=\frac{\Pi ^{\prime }(M^{2},s_{0})}{\Pi (M^{2},s_{0})},  \label{eq:Mass}
\end{equation}%
and
\begin{equation}
\Lambda ^{2}=e^{m^{2}/M^{2}}\Pi (M^{2},s_{0}).  \label{eq:Coupl}
\end{equation}%
In Eqs.\ (\ref{eq:Mass}) and (\ref{eq:Coupl}) $\Pi (M^{2},s_{0})$ is the
amplitude $\Pi ^{\mathrm{OPE}}(p^{2})$ after the Borel transformation and
continuum subtraction, and $\Pi ^{\prime }(M^{2},s_{0})=d\Pi
(M^{2},s_{0})/d(-1/M^{2})$. The Borel transformation is applied to suppress
contributions of higher resonances and continuum states, while using the
continuum subtraction we remove them from the QCD side of the SR equality.

These manipulations generates a dependence of $\Pi (M^{2},s_{0})$ on the
Borel $M^{2}$ and continuum subtraction $s_{0}$ parameters. It is given by
the expression%
\begin{equation}
\Pi (M^{2},s_{0})=\int_{(3m_{c}+m_{b})^{2}}^{s_{0}}ds\rho ^{\mathrm{OPE}%
}(s)e^{-s/M^{2}}+\Pi (M^{2}),
\end{equation}%
where $\rho ^{\mathrm{OPE}}(s)$ is the spectral density which is equal to an
imaginary part of $\Pi ^{\mathrm{OPE}}(p^{2})$. Because we take into account
contributions to $\Pi ^{\mathrm{OPE}}(p^{2})$ arising from the perturbative
and terms proportional to $\langle \alpha _{s}G^{2}/\pi \rangle $ , $\rho ^{%
\mathrm{OPE}}(s)$ is a sum of the components $\rho ^{\mathrm{pert.}}(s)$ and
$\rho ^{\mathrm{Dim4}}(s)$. The function $\Pi (M^{2})$ does not contain
contributions included into $\rho ^{\mathrm{OPE}}(s)$ and is derived from
the correlation function $\Pi ^{\mathrm{OPE}}(p)$ .

To start numerical computations we fix the gluon condensate $\langle \alpha
_{s}G^{2}/\pi \rangle =(0.012\pm 0.004)~\mathrm{GeV}^{4}$ and masses $%
m_{c}=(1.2730\pm 0.0046)~\mathrm{GeV}$ and $m_{b}=(4.183\pm 0.007)~\mathrm{%
GeV}$ of the quarks which are universal quantities \cite%
{Shifman:1978bx,Shifman:1978by,PDG:2024}. The parameters $M^{2}$ and $s_{0}$
depend on a considering process and have to satisfy usual constraints of SR
studies. The prevalence of the pole contribution ($\mathrm{PC}$) in
extracted quantities is one of such important restrictions. Therefore, we
choose $M^{2}$ and $s_{0}$ in such a way that to ensure $\mathrm{PC}\geq 0.5$
. The next constraint is convergence of the operator product expansion. We
calculate $\Pi (M^{2},s_{0})$ by taking into account dimension-$4$
nonperturbative terms, therefore require fulfillment of $|\Pi ^{\mathrm{Dim4}%
}(M^{2},s_{0})|\leq 0.05|\Pi (M^{2},s_{0})|$ which is sufficient for convergence
of $\mathrm{OPE}$. Additionally, physical parameters of $\mathfrak{M}$ have
to be stable upon variations of $M^{2}$ and $s_{0}$.

\begin{figure}[h]
\includegraphics[width=8.5cm]{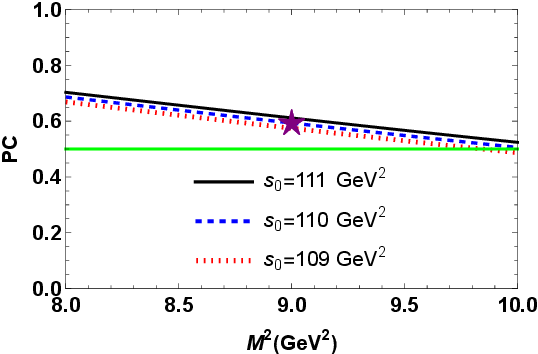}
\caption{$\mathrm{PC}$ as a function of the Borel parameter $M^{2}$ at fixed
$s_{0}$. The star shows the point $M^{2}=9~\mathrm{GeV}^{2},\ s_{0}=110~%
\mathrm{GeV}^{2}$. }
\label{fig:PC}
\end{figure}

We have conducted numerical analysis of $m$ and $\Lambda $ using broad
windows for the parameters $M^{2}$ and $s_{0}$. Information gathered during
these computations has permitted us to restrict limits of $M^{2}$ and $s_{0}$
\begin{equation}
M^{2}\in \lbrack 8,10]~\mathrm{GeV}^{2},\ s_{0}\in \lbrack 109,111]~\mathrm{%
GeV}^{2},  \label{eq:Wind1}
\end{equation}%
where all aforementioned constraints are satisfied. Thus, the $s_{0}$%
-averaged $\mathrm{PC}$ at $M^{2}=10~\mathrm{GeV}^{2}$ is equal to $0.52$,
meanwhile at $M^{2}=8~\mathrm{GeV}$ it amounts to $0.78$. At $M^{2}=8~%
\mathrm{GeV}^{2}$ the term $\Pi ^{\mathrm{Dim4}}(M^{2},s_{0})$ and does not
exceed $2\%$ of $\Pi (M^{2},s_{0})$. Dependence of the pole contribution on $%
M^{2}$ is plotted in Fig.\ \ref{fig:PC}.

\begin{figure}[h!]
\begin{center}
\includegraphics[totalheight=6cm,width=8cm]{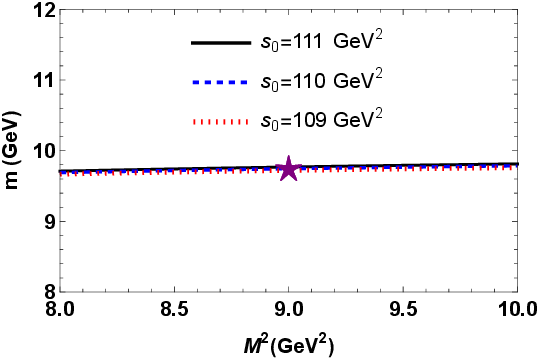} %
\includegraphics[totalheight=6cm,width=8cm]{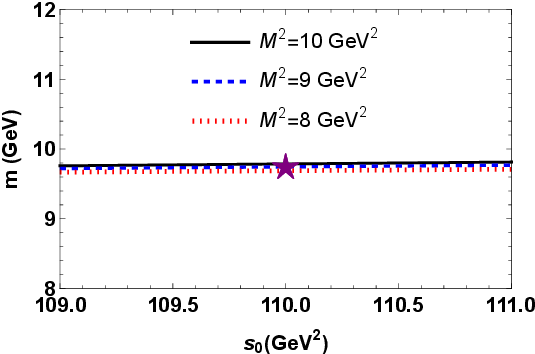}
\end{center}
\caption{Dependence of $m$ on the parameters $M^{2}$ (left panel), and $s_0$
(right panel).}
\label{fig:Mass}
\end{figure}

The parameters $m$ and $\Lambda $ extracted using SRs and regions Eq.\ (\ref%
{eq:Wind1}) are equal to
\begin{equation}
m=(9740\pm 70)~\mathrm{MeV},\ \ \Lambda =(5.19\pm 0.49)~\times 10^{-1}\text{
}\mathrm{GeV}^{5}.  \label{eq:Result1}
\end{equation}%
Ambiguities in Eq.\ (\ref{eq:Result1}) amount to $\pm 0.7\%$ for $m$, and to
$\pm 9.5\%$ in the case of the parameter $\Lambda $. They are mostly
connected with the choice $M^{2}$ and $s_{0}$. The masses of quarks and
gluon condensate also contains errors, but their effects on the results are
negligible. It is worth noting that uncertainties in Eq.\ (\ref{eq:Result1})
are acceptable for SR computations. The mass $m$ as a function of $M^{2}$
and $s_{0}$ is depicted in Fig.\ \ref{fig:Mass}. One sees that lines
presented in this figure are stable (within existing errors) upon variations
of the parameters $M^{2}$ and $s_{0}$.

The spectroscopic parameters $\widetilde{m}$ and $\widetilde{\Lambda }$ of
the molecule $\widetilde{\mathfrak{M}}$ are calculated in accordance with a
scheme described above. Computations lead to the following predictions
\begin{equation}
\widetilde{m}=(9737\pm 71)~\mathrm{MeV},\ \ \widetilde{\Lambda }=(5.23\pm
0.50)~\times 10^{-1}\text{ }\mathrm{GeV}^{5}.  \label{eq:Result2}
\end{equation}%
It is clear that these parameters are very close to ones written down in
Eq.\ (\ref{eq:Result1}): There are wide overlapping regions for $m$, $%
\Lambda $ and $\widetilde{m}$, $\widetilde{\Lambda }$. Because theoretical
errors do not allow us to distinguish clearly the states $\mathfrak{M}$ and $%
\widetilde{\mathfrak{M}}$, in what follows, we explore the molecule $%
\mathfrak{M}$.


\section{Dominant decay channels}

\label{sec:Widths1}

This section is devoted to analysis of the dominant decays $\mathfrak{%
M\rightarrow }J/\psi B_{c}^{+}$ and $\mathfrak{M\rightarrow }\eta
_{c}B_{c}^{\ast +}$ of the hadronic molecule $\mathfrak{M}$. We calculate
the strong couplings $g_{1}$ and $g_{2}$ at the vertices $\mathfrak{M}J/\psi
B_{c}^{+}$ and $\mathfrak{M}\eta _{c}B_{c}^{\ast +}$, respectively: They are
necessary to find the partial widths of these processes. To this end, we
invoke technical tools of the three-point SR method and evaluate the form
factors $g_{1}(q^{2})$ and $g_{2}(q^{2})$. At the mass shells of the mesons $%
J/\psi $ and $\eta _{c}$ the form factors $g_{1}(m_{J/\psi }^{2})$ and $%
g_{2}(\eta _{c}^{2})$ give the required strong couplings $g_{1}$ and $g_{2}$.


\subsection{$\mathfrak{M\rightarrow }J/\protect\psi B_{c}^{+}$}


The three-point correlation function to extract the sum rule for the form
factor $g_{1}(q^{2})$ is given by the expression
\begin{equation}
\Pi _{\mu \nu }^{1}(p,p^{\prime })=i^{2}\int d^{4}xd^{4}ye^{ip^{\prime
}y}e^{-ipx}\langle 0|\mathcal{T}\{I^{B_{c}}(y)I_{\mu }^{J/\psi }(0)I_{\nu
}^{\dagger }(x)\}|0\rangle ,  \label{eq:CF1a}
\end{equation}%
where $I^{B_{c}}(x)$ and $I_{\mu }^{J/\psi }$ are currents that correspond
to mesons $B_{c}^{+}$ and $J/\psi $, respectively. They are introduced by
the formulas
\begin{equation}
I_{\mu }^{J/\psi }(x)=\overline{c}_{i}(x)\gamma _{\mu }c_{i}(x),I^{B_{c}}(x)=%
\overline{b}_{j}(x)i\gamma _{5}c_{j}(x).
\end{equation}

The correlator $\Pi _{\mu \nu }^{1}(p,p^{\prime })$ written down using
masses and decay constants (current coupling) of the particles $\mathfrak{M}$%
, $J/\psi $, and $B_{c}^{+}$ forms the phenomenological side $\Pi _{\mu \nu
}^{1\mathrm{Phys}}(p,p^{\prime })$ of SR. After dissecting the contribution
of the ground-state particles in the factorization approximation \cite%
{Ioffe:1982qb,Colangelo:2000dp,Agaev:2022iha}, it has the form
\begin{eqnarray}
\Pi _{\mu \nu }^{1\mathrm{Phys}}(p,p^{\prime }) &=&\frac{\langle
0|I^{B_{c}}|B_{c}^{+}(p^{\prime })\rangle }{p^{\prime 2}-m_{B_{c}}^{2}}\frac{%
\langle 0|I_{\mu }^{J/\psi }|J/\psi (q,\varepsilon )\rangle }{%
q^{2}-m_{J/\psi }^{2}}\langle J/\psi (q,\varepsilon )B_{c}^{+}(p^{\prime })|%
\mathfrak{M}(p,\epsilon )\rangle  \notag \\
&&\times \frac{\langle \mathfrak{M}(p,\epsilon )|I_{\nu }^{\dagger
}|0\rangle }{p^{2}-m^{2}}+\cdots ,  \label{eq:TP1}
\end{eqnarray}%
where $m_{B_{c}}=(6274.47\pm 0.27\pm 0.17)~\mathrm{MeV}$ and $m_{J/\psi
}=(3096.900\pm 0.006)~\mathrm{MeV}$ are the masses of the mesons $B_{c}^{+}$
and $J/\psi $ \cite{PDG:2024}, while $\varepsilon _{\mu }$ is the
polarization vector of the charmonium $J/\psi $.

After applying the matrix elements of the mesons $B_{c}^{+}$ and $J/\psi $
\begin{equation}
\langle 0|I_{\mu }^{J/\psi }|J/\psi (q,\varepsilon )\rangle =f_{J/\psi
}m_{J/\psi }\varepsilon _{\mu },\ \ \langle 0|I^{B_{c}}|B_{c}^{+}(p^{\prime
})\rangle =\frac{f_{B_{c}}m_{B_{c}}^{2}}{m_{b}+m_{c}},  \label{eq:ME1A}
\end{equation}%
as well as the matrix element for the vertex
\begin{equation}
\langle J/\psi (q,\varepsilon )B_{c}^{+}(p^{\prime })|\mathfrak{M}%
(p,\epsilon )\rangle =g_{1}(q^{2})\left[ (p\cdot p^{\prime })(\epsilon \cdot
\varepsilon ^{\ast })-(p^{\prime }\cdot \epsilon )(p\cdot \varepsilon ^{\ast
})\right] .  \label{eq:AVVPS}
\end{equation}%
we can express $\Pi _{\mu \nu }^{1\mathrm{Phys}}(p,p^{\prime })$ using the
physical parameters of $\mathfrak{M}$, $J/\psi $, and $B_{c}^{+}$ involved
into decay process. As a result, we find
\begin{eqnarray}
&&\Pi _{\mu \nu }^{1\mathrm{Phys}}(p,p^{\prime })=\frac{g_{1}(q^{2})\Lambda
f_{B_{c}}m_{B_{c}}^{2}f_{J/\psi }m_{J/\psi }}{(m_{b}+m_{c})\left(
p^{2}-m^{2}\right) (p^{\prime 2}-m_{B_{c}}^{2})}\frac{1}{(q^{2}-m_{J/\psi
}^{2})}\left[ \frac{m^{2}+m_{B_{c}}^{2}-q^{2}}{2}g_{\mu \nu }\right.  \notag
\\
&&\left. -\frac{m^{2}}{m_{J/\psi }^{2}}p_{\mu }^{\prime }p_{\nu }^{\prime }+%
\frac{m^{2}-m_{J/\psi }^{2}}{m_{J/\psi }^{2}}p_{\mu }^{\prime }p_{\nu }-%
\frac{m^{2}+m_{B_{c}}^{2}-q^{2}}{2m_{J/\psi }^{2}}(p_{\mu }p_{\nu }+p_{\mu
}p_{\nu }^{\prime })\right] +\cdots .  \label{eq:PhysSide1}
\end{eqnarray}%
In Eqs.\ (\ref{eq:ME1A}) and (\ref{eq:PhysSide1}) $f_{J/\psi }=(411\pm 7)~%
\mathrm{MeV}$ and $\ f_{B_{c}}=(371\pm 37)~\mathrm{MeV}$ are the decay
constants of $J/\psi $ and $B_{c}^{+}$ \cite{Lakhina:2006vg,Wang:2024fwc},
respectively.

The QCD side of SR is determined by the correlator $\Pi _{\mu \nu }^{1%
\mathrm{OPE}}(p,p^{\prime })$
\begin{eqnarray}
&&\Pi _{\mu \nu }^{1\mathrm{OPE}}(p,p^{\prime })=-\int
d^{4}xd^{4}ye^{ip^{\prime }y}e^{-ipx}\left\{ \mathrm{Tr}\left[ \gamma _{\mu
}S_{c}^{ja}(-x)\gamma _{\nu }S_{c}^{aj}(x)\right] \mathrm{Tr}\left[ \gamma
_{5}S_{c}^{ib}(y-x)\gamma _{5}S_{b}^{bi}(x-y)\right] \right.  \notag \\
&&\left. -\mathrm{Tr}\left[ \gamma _{5}S_{c}^{ia}(y-x)\gamma _{\nu
}S_{c}^{aj}(x)\gamma _{\mu }S_{c}^{jb}(-x)\gamma _{5}S_{b}^{bi}(x-y)\right]
\right\} .  \label{eq:CF3}
\end{eqnarray}%
The correlation functions $\Pi _{\mu \nu }^{1\mathrm{Phys}}(p,p^{\prime })$
and $\Pi _{\mu \nu }^{1\mathrm{OPE}}(p,p^{\prime })$ contains different
Lorentz structures. We choose to work with the invariant amplitudes $\Pi
_{1}^{\mathrm{Phys}}(p^{2},p^{\prime 2},q^{2})$ and $\Pi _{1}^{\mathrm{OPE}%
}(p^{2},p^{\prime 2},q^{2})$ which are related to terms $\sim g_{\mu \nu }$.

Having equated the functions $\Pi _{1}^{\mathrm{Phys}}(p^{2},p^{\prime
2},q^{2})$ and $\Pi _{1}^{\mathrm{OPE}}(p^{2},p^{\prime 2},q^{2})$,
performed\ the double Borel transformations (over the variables $-p^{2}$, $%
-p^{\prime 2}$), and subtracted contributions of suppressed terms, we get SR
for $g_{1}(q^{2})$
\begin{equation}
g_{1}(q^{2})=\frac{2(m_{b}+m_{c})}{\Lambda f_{B_{c}}m_{B_{c}}^{2}f_{J/\psi
}m_{J/\psi }}\frac{q^{2}-m_{J/\psi }^{2}}{m^{2}+m_{B_{c}}^{2}-q^{2}}%
e^{m^{2}/M_{1}^{2}}e^{m_{B_{c}}^{2}/M_{2}^{2}}\Pi _{1}(\mathbf{M}^{2},%
\mathbf{s}_{0},q^{2}).  \label{eq:SRCoupl1}
\end{equation}%
Here, $\Pi _{1}(\mathbf{M}^{2},\mathbf{s}_{0},q^{2})$ is the amplitude $\Pi
_{1}^{\mathrm{OPE}}(p^{2},p^{\prime 2},q^{2})$ after the double Borel
transformation and continuum subtractions. It is given by the formula
\begin{equation}
\Pi _{1}(\mathbf{M}^{2},\mathbf{s}_{0},q^{2})=%
\int_{(3m_{c}+m_{b})^{2}}^{s_{0}}ds\int_{(m_{b}+m_{c})^{2}}^{s_{0}^{\prime
}}ds^{\prime }\rho _{1}(s,s^{\prime },q^{2})e^{-s/M_{1}^{2}}e^{-s^{\prime
}/M_{2}^{2}},  \label{eq:AS1}
\end{equation}%
where spectral density $\rho (s,s^{\prime },q^{2})$ is determined as the
imaginary part of $\Pi _{1}^{\mathrm{OPE}}(s,s^{\prime },q^{2})$.

The function $\Pi _{1}(\mathbf{M}^{2},\mathbf{s}_{0},q^{2})$ depends on two
pairs of parameters $(M_{1}^{2},s_{0})$ and $(M_{2}^{2},s_{0}^{\prime })$.
The first of them corresponds to the channel of the molecule $\mathfrak{M}$,
while $(M_{2}^{2},s_{0}^{\prime })$ is related to the channel of $B_{c}^{+}$
meson. They are restricted in accord with standard rules of SR calculations.
Our studies prove that Eq.\ (\ref{eq:Wind1}) for the parameters $%
(M_{1}^{2},s_{0})$ and 
\begin{equation}
M_{2}^{2}\in \lbrack 6.5,7.5]~\mathrm{GeV}^{2},\ s_{0}^{\prime }\in \lbrack
45,47]~\mathrm{GeV}^{2}.  \label{eq:Wind3}
\end{equation}%
meet SR restrictions.

\begin{figure}[h!]
\begin{center}
\includegraphics[totalheight=6cm,width=8cm]{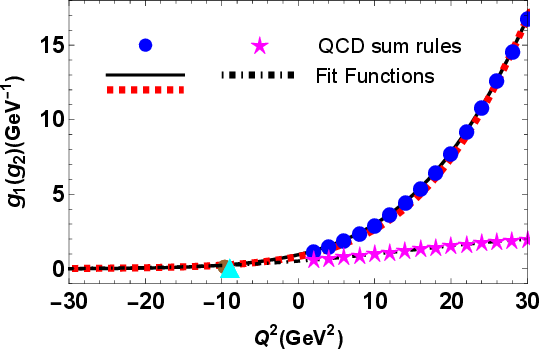} %
\includegraphics[totalheight=6cm,width=8cm]{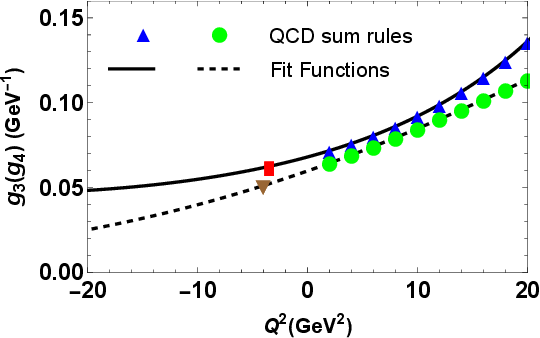}
\end{center}
\caption{Left panel: SR data and fit functions $\mathcal{Z}_1 (Q^{2})$
(solid curve), $\mathcal{Z}_{1A} (Q^{2})$ (dashed line) and $\mathcal{Z}_2
(Q^{2})$ (dot-dashed line). The triangle and circle denote values of the
extrapolating functions at the points $Q^{2}=-m_{J/\protect\psi}^{2}$ and $%
Q^{2}=-m_{\protect\eta_c}^2$, respectively. Right panel: QCD data and
extrapolating functions $\mathcal{Z}_3 (Q^{2})$ (solid) and $\mathcal{Z}_4
(Q^{2})$ (dashed). The rectangle and triangle fix fit functions at the
points $Q^{2}=-m_{D^{0}}^{2}$ and $Q^{2}=-m_{D^{\ast 0}}^2$. }
\label{fig:Fit}
\end{figure}

It is known that the SR method leads to credible results in the Euclidean
region $q^{2}<0$, whereas $g_{1}(q^{2})$ amounts to $g_{1}$ at $q^{2}=m_{J/\psi
}^{2}$. Hence, it is convenient to make use of new function $g_{1}(Q^{2})$
where $Q^{2}=-q^{2}$, and employ it for subsequent investigations. The QCD
results for $g_{1}(Q^{2})$ where $Q^{2}=2-30~\mathrm{GeV}^{2}$are plotted in
Fig.\ \ref{fig:Fit}.

To extract $g_{1}$ at $q^{2}=-Q^{2}=m_{J/\psi }^{2}$, we employ the
extrapolating function $\mathcal{Z}_{1}(Q^{2})$: At $Q^{2}>0$ it is equal to
the SR data, but can be applied in the region of $Q^{2}<0$. We fix $\mathcal{%
Z}_{1}(Q^{2})$ in the following form
\begin{equation}
\mathcal{Z}_{i}(Q^{2})=\mathcal{Z}_{i}^{0}\mathrm{\exp }\left[ z_{i}^{1}%
\frac{Q^{2}}{m^{2}}+z_{i}^{2}\left( \frac{Q^{2}}{m^{2}}\right) ^{2}\right] ,
\label{eq:FitF}
\end{equation}%
where the parameters $\mathcal{Z}_{i}^{0}$, $z_{i}^{1}$, and $z_{i}^{2}$ are
estimated from comparison of $\mathcal{Z}_{1}(Q^{2})$ and SR data. It is not
difficult to find that
\begin{equation}
\mathcal{Z}_{1}^{0}=0.945~\mathrm{GeV}^{-1},~z_{1}^{1}=11.85,\text{~}%
z_{1}^{2}=-8.66.  \label{eq:FF1}
\end{equation}%
In Fig.\ \ref{fig:Fit} we plot also $\mathcal{Z}_{1}(Q^{2})$, in which nice
agreement of $\mathcal{Z}_{1}(Q^{2})$ and SR data is evident. Then, for $%
g_{1}$ we get
\begin{equation}
g_{1}\equiv \mathcal{Z}_{1}(-m_{J/\psi }^{2})=(2.61\pm 0.50)\times 10^{-1}\
\mathrm{GeV}^{-1}.  \label{eq:g1}
\end{equation}

This result has been found by using Eq.\ (\ref{eq:FitF}). At the same time,
SR data can be extended into region the region $Q^{2}<0$ by employing
different extrapolating functions. To clarify this problem, we examine the
fit function
\begin{equation}
\mathcal{Z}_{1A}(Q^{2})=\frac{c_{0}}{\left( 1-Q^{2}/m^{2}\right)
^{4}[1-c_{1}(Q^{2}/m^{2})+c_{2}\left( Q^{4}/m^{4}\right) ]},
\label{eq:FitF2}
\end{equation}%
where $c_{0}$, $c_{1}$, and $c_{2}$ are fitted constants. We get $%
c_{0}=0.826\ \mathrm{GeV}^{-1}$, $c_{1}=1.844$, and $c_{2}=2.731$ from
comparison of Eq. (\ref{eq:FitF2}) and SR data. This function is presented
in Fig.\ \ref{fig:Fit} as well, in which one sees that it agrees with SR
data and $\mathcal{Z}_{1}(Q^{2})$. The fit function $\mathcal{Z}_{1A}(Q^{2})$
leads to the result $g_{1}=0.158~\mathrm{GeV}^{-1}$, which deviates $|0.003|$
from the value in Eq.\ (\ref{eq:g1}). \ But this deviation is more than an
order of magnitude smaller than uncertainties $\pm 0.05$ of $g_{1}$ arising
from the sum rule method. For this reason, we neglect uncertainties
appearing due to the choice of extrapolating functions and employ Eq.\ (\ref%
{eq:FitF}) in our computations. It is worth noting that analytical forms of
the extrapolating functions $\mathcal{Z}_{1}(Q^{2})$ and $\mathcal{Z}%
_{1A}(Q^{2})$ are inspired by the SR for $g_{1}(q^{2})$ in Eq.\ (\ref%
{eq:SRCoupl1}), which contains both the exponential and $(m^{2}-q^{2})$-type
factors.

The width of the decay $\mathfrak{M\rightarrow }J/\psi B_{c}^{+}$ is
computed using the formula
\begin{equation}
\Gamma \left[ \mathfrak{M\rightarrow }J/\psi B_{c}^{+}\right] =g_{1}^{2}%
\frac{\lambda _{1}}{24\pi m^{2}}|M_{1}|^{2}.  \label{eq:PDw1}
\end{equation}%
Here,
\begin{equation}
|M_{1}|^{2}=\frac{1}{4m_{B_{c}}^{2}}\left[ m^{6}-2m^{4}m_{J/\psi
}^{2}+2m_{B_{c}}^{2}(m_{B_{c}}^{2}-m_{J/\psi }^{2})^{2}+m^{2}(m_{J/\psi
}^{4}+6m_{J/\psi }^{2}m_{B_{c}}^{2}-3m_{B_{c}}^{4})\right] .
\end{equation}%
We have also employed $\lambda _{1}=\lambda (m,m_{B_{c}},m_{J/\psi })$ where
$\lambda (a,b,c)$ is
\begin{equation}
\lambda (a,b,c)=\frac{\sqrt{%
a^{4}+b^{4}+c^{4}-2(a^{2}b^{2}+a^{2}c^{2}+b^{2}c^{2})}}{2a}.
\end{equation}

As a result, one finds
\begin{equation}
\Gamma \left[ \mathfrak{M\rightarrow }J/\psi B_{c}^{+}\right] =(40.6\pm
12.4)~\mathrm{MeV}.  \label{eq:DW2}
\end{equation}

\subsection{$\mathfrak{M\rightarrow }\protect\eta _{c}B_{c}^{\ast +}$}


To explore the channel $\mathfrak{M\rightarrow }\eta _{c}B_{c}^{\ast +}$ we
estimate the coupling $g_{2}$ at the vertex $\mathfrak{M}\eta
_{c}B_{c}^{\ast +}$. To this end, it is convenient to consider the
correlation function $\Pi _{\mu \nu }^{2}(p,p^{\prime })$
\begin{equation}
\Pi _{\mu \nu }^{2}(p,p^{\prime })=i^{2}\int d^{4}xd^{4}ye^{ip^{\prime
}y}e^{-ipx}\langle 0|\mathcal{T}\{I_{\mu }^{B_{c}^{\ast }}(y)I^{\eta
_{c}}(0)I_{\nu }^{\dagger }(x)\}|0\rangle .  \label{eq:CF1b}
\end{equation}%
In Eq.\ (\ref{eq:CF1b}) $I^{\eta _{c}}(x)$ and $I_{\mu }^{B_{c}^{\ast }}(x)$
are the interpolating currents for the mesons $\eta _{c}$ and $B_{c}^{\ast
+} $
\begin{equation}
I^{\eta _{c}}(x)=\overline{c}_{i}(x)i\gamma _{5}c_{i}(x),\ I_{\mu
}^{B_{c}^{\ast }}(x)=\overline{b}_{j}(x)\gamma _{\mu }c_{j}(x).
\end{equation}%
The correlation function $\Pi _{\mu \nu }^{2\mathrm{Phys}}(p,p^{\prime })$
has the form%
\begin{eqnarray}
\Pi _{\mu \nu }^{2\mathrm{Phys}}(p,p^{\prime })&=&\frac{g_{2}(q^{2})\Lambda
f_{\eta _{c}}m_{\eta _{c}}^{2}f_{B_{c}^{\ast }}m_{B_{c}^{\ast }}}{%
2m_{c}\left( p^{2}-m^{2}\right) \left( p^{\prime 2}-m_{B_{c}^{\ast
}}^{2}\right) (q^{2}-m_{\eta _{c}}^{2})}\left[ \frac{m^{2}-m_{B_{c}^{\ast
}}^{2}+q^{2}}{2}g_{\mu \nu }-p_{\mu }p_{\nu }+p_{\mu }^{\prime }p_{\nu
}\right.  \notag \\
&&\left. +\frac{m^{2}+m_{B_{c}^{\ast }}^{2}-q^{2}}{m_{B_{c}^{\ast }}^{2}}%
p_{\mu }p_{\nu }^{\prime }\right] +\cdots .
\end{eqnarray}%
To find $\Pi _{\mu \nu }^{2\mathrm{Phys}}(p,p^{\prime })$ the matrix
elements
\begin{equation}
\langle 0|I_{\mu }^{B_{c}^{\ast }}|B_{c}^{\ast +}(p^{\prime },\varepsilon
)\rangle =f_{B_{c}^{\ast }}m_{B_{c}^{\ast }}\varepsilon _{\mu },\ \ \langle
0|I^{\eta _{b}}|\eta _{c}(q)\rangle =\frac{f_{\eta _{c}}m_{\eta _{c}}^{2}}{%
2m_{c}},
\end{equation}%
and
\begin{equation}
\langle \eta _{c}(q)B_{c}^{\ast +}(p^{\prime },\varepsilon )|\mathfrak{M}%
(p,\epsilon )\rangle =g_{2}(q^{2})\left[ (p\cdot q)(\epsilon \cdot
\varepsilon ^{\ast })-(q\cdot \epsilon )(p\cdot \varepsilon ^{\ast })\right]
,
\end{equation}%
have been used. Above, $m_{\eta _{c}}=$ $(2984.1\pm 0.4)~\mathrm{MeV}$ and $%
f_{\eta _{c}}=(421\pm 35)~\mathrm{MeV}$ are parameters of the charmonium $%
\eta _{c}$. The mass $m_{B_{c}^{\ast }}=$ $6338~\mathrm{MeV}$ and $%
f_{B_{c}^{\ast }}=471~\mathrm{MeV}$ are model predictions obtained in Refs.\
\cite{Godfrey:2004ya,Eichten:2019gig}.

The QCD side of the SR is given by the formula
\begin{equation}
\Pi _{\mu \nu }^{2\mathrm{OPE}}(p,p^{\prime })=\int
d^{4}xd^{4}ye^{ip^{\prime }y}e^{-ipx}\mathrm{Tr}\left[ \gamma _{\mu
}S_{c}^{ia}(y-x)\gamma _{\nu }S_{c}^{aj}(x)\gamma _{5}S_{c}^{jb}(-x)\gamma
_{5}S_{b}^{bi}(x-y)\right] .
\end{equation}%
To proceed we use the invariant amplitudes $\Pi _{2}^{\mathrm{Phys}%
}(p^{2},p^{\prime 2},q^{2})$ and $\Pi _{2}^{\mathrm{OPE}}(p^{2},p^{\prime
2},q^{2})$ related to structures $\sim g_{\mu \nu }$ both in the physical
and $\mathrm{OPE}$ versions of the correlator $\Pi _{\mu \nu
}^{2}(p,p^{\prime })$. The SR for $g_{2}(q^{2})$ derived after standard
manipulations reads
\begin{equation}
g_{2}(q^{2})=\frac{4m_{c}}{\Lambda f_{\eta _{c}}m_{\eta
_{c}}^{2}f_{B_{c}^{\ast }}m_{B_{c}^{\ast }}}\frac{q^{2}-m_{\eta _{c}}^{2}}{%
m^{2}-m_{B_{c}^{\ast }}^{2}+q^{2}}e^{m^{2}/M_{1}^{2}}e^{m_{B_{c}^{\ast
}}^{2}/M_{2}^{2}}\Pi _{2}(\mathbf{M}^{2},\mathbf{s}_{0},q^{2}),
\end{equation}%
respectively. The regions
\begin{equation}
M_{2}^{2}\in \lbrack 6.5,7.5]~\mathrm{GeV}^{2},\ s_{0}^{\prime }\in \lbrack
49,51]~\mathrm{GeV}^{2},
\end{equation}%
in $B_{c}^{\ast +}$ meson's channel satisfy required constraints of SR
analysis. The corresponding fit function $\mathcal{Z}_{2}(Q^{2})$ has the
parameters
\begin{equation}
\mathcal{Z}_{2}^{0}=0.518~\mathrm{GeV}^{-1},\ z_{2}^{1}=7.07,\
z_{2}^{2}=-8.93.
\end{equation}%
The SR data and the function $\mathcal{Z}_{2}(Q^{2})$ are also shown in
Fig.\ \ref{fig:Fit}. For the strong coupling $g_{2}$, we find
\begin{equation}
g_{2}\equiv \mathcal{Z}_{2}(-m_{B_{c}^{\ast }}^{2})=(2.5\pm 0.4)\times
10^{-1}\ \mathrm{GeV}^{-1}.
\end{equation}%
The width of the channel $\mathfrak{M\rightarrow }\eta _{c}B_{c}^{\ast +}$
is
\begin{equation}
\Gamma \left[ \mathfrak{M\rightarrow }\eta _{c}B_{c}^{\ast +}\right]
=(36.8\pm 9.5)~\mathrm{MeV}.
\end{equation}


\section{Subdominant decays $\mathfrak{M\rightarrow }B^{\ast +}D^{0}$, $%
B^{\ast 0}D^{+}$, $B^{+}D^{\ast 0}$, and $B^{0}D^{\ast +}$}

\label{sec:Widths2}


Annihilation of $\overline{c}c$ quarks generates numerous subdominant decay
channels of the molecule $J/\psi B_{c}^{+}$. Processes $\mathfrak{%
M\rightarrow }B^{\ast +}D^{0}$, $B^{\ast 0}D^{+}$, $B^{+}D^{\ast 0}$, and $%
B^{0}D^{\ast +}$are such decay modes. Importance of this mechanism to
explore decays of exotic states was emphasized in Refs.\ \cite%
{Becchi:2020mjz,Becchi:2020uvq,Agaev:2023ara} and used to explore decays of
numerous tetraquarks. It was also applied to investigate processes with
hadronic molecules \cite%
{Agaev:2025wdj,Agaev:2025fwm,Agaev:2025nkw,Agaev:2025did}

It is worth emphasizing that some of these decays have parameters very close
to each other. The reason is that in the limit $m_{\mathrm{u}}=m_{\mathrm{d}%
}=0$, which is adopted in this work, phenomenological components of the
relevant SRs for the decays $\mathfrak{M\rightarrow }B^{\ast +}D^{0}$ and $%
B^{\ast 0}D^{+}$ differ by the masses and decay constants of, for instance, $%
D^{0}$ and $D^{+}$ mesons. The QCD components of the sum rules contain $u$
and $d$ quarks' propagators identical in this case. Because we ignore small
differences in the masses and decay constants of the mesons, the SRs lead to
the same predictions of their parameters. Hence, we treat these decays as
identical processes and calculate the partial width of the channel $%
\mathfrak{M\rightarrow }B^{\ast +}D^{0}$: The width of the second mode is
equal to that of the first one. These arguments are also valid for the pair
of processes $\mathfrak{M\rightarrow }B^{+}D^{\ast 0}$ and $\mathfrak{%
M\rightarrow }B^{0}D^{\ast +}$.


\subsection{$\mathfrak{M\rightarrow }B^{\ast +}D^{0}$ and $B^{\ast 0}D^{+}$}


Here, we study  the mode $\mathfrak{M\rightarrow }B^{\ast +}D^{0}$
and calculate its partial decay width. We plan to determine the strong
coupling $g_{3}$ of particles at $\mathfrak{M}B^{\ast +}D^{0}$. The
correlation function to be considered is
\begin{equation}
\Pi _{\mu \nu }^{3}(p,p^{\prime })=i^{2}\int d^{4}xd^{4}ye^{ip^{\prime
}y}e^{-ipx}\langle 0|\mathcal{T}\{I_{\mu }^{B^{\ast }}(y)I^{D^{0}}(0)I_{\nu
}^{\dagger }(x)\}|0\rangle ,  \label{eq:CF3A}
\end{equation}%
where $I_{\mu }^{B^{\ast }}$ and $I^{D^{0}}(x)$ are currents for the
particles $B^{\ast -}$ and $D^{0}$ mesons
\begin{equation}
I_{\mu }^{B^{\ast }}(x)=\overline{b}_{i}(x)\gamma _{\mu }u_{i}(x),\
I^{D^{0}}(x)=\overline{u}_{j}(x)i\gamma _{5}c_{j}(x).
\end{equation}

The phenomenological formula for $\Pi _{\mu \nu }^{3}(p,p^{\prime })$ is
found by means of the expressions%
\begin{equation}
\langle 0|I_{\mu }^{B^{\ast }}|B^{\ast +}(p^{\prime },\varepsilon )\rangle
=f_{B^{\ast }}m_{B^{\ast }}\varepsilon _{\mu },\ \ \ \langle
0|I^{D^{0}}|D^{0}(q)\rangle =\frac{f_{D}m_{D^{0}}^{2}}{m_{c}}.
\label{eq:ME3}
\end{equation}%
The characteristics of the particles above are $m_{B^{\ast }}=(5324.75\pm
0.20)~\mathrm{MeV},\ m_{D^{0}}=(1864.84\pm 0.05)~\mathrm{MeV}$ and $%
f_{B^{\ast }}=(210\pm 6)~\mathrm{MeV},$ $f_{D}=(211.9\pm 1.1)~\mathrm{MeV}$.
In Eq.\ (\ref{eq:ME3}) the polarization vector of the vector meson $B^{\ast
+}$ is denoted by $\varepsilon _{\mu }$. The vertex $\langle B^{\ast
+}(p^{\prime },\varepsilon )D^{0}(q)|\mathfrak{M}(p,\epsilon )\rangle $ and $%
\Pi _{\mu \nu }^{3\mathrm{Phys}}(p,p^{\prime })$ are similar to expressions
presented in the previous section \ref{sec:Widths1}.

Computations of $\Pi _{\mu \nu }^{3}(p,p^{\prime })$ yield
\begin{equation}
\Pi _{\mu \nu }^{3\mathrm{OPE}}(p,p^{\prime })=\frac{1}{3}\int
d^{4}xd^{4}ye^{ip^{\prime }y}e^{-ipx}\langle \overline{c}c\rangle \mathrm{Tr}%
\left[ \gamma _{\mu }S_{u}^{ij}(y)\gamma _{5}S_{c}^{ja}(-x)\gamma _{\nu
}\gamma _{5}S_{b}^{ia}(x-y)\right] .
\end{equation}%
The correlation function $\Pi _{\mu \nu }^{3\mathrm{OPE}}(p,p^{\prime })$
depends on three quark propagators and vacuum condensate $\langle \overline{c%
}c\rangle $, and differs from standard correlators which contain four quark
propagators. In our computations we contract heavy and light quark fields.
The mesons $B^{\ast +}D^{0}$ contain only one $c$-quark field, as a result,
free $\overline{c}c$ quarks in $\mathfrak{M}$ form a local vacuum
condensate. In other words, $\langle \overline{c}c\rangle $ appears in
formulas instead of the quark propagator and is considering on equal footing
with them.

In what follows, we apply the relation between the heavy quark and gluon
condensates to express $\Pi _{\mu \nu }^{3\mathrm{OPE}}(p,p^{\prime })$
using relevant parameters. The available formula for $\langle \overline{c}%
c\rangle $ reads \cite{Shifman:1978bx,Generalis:1983hb,Bagan:1985zp}
\begin{equation}
m_{c}\langle \overline{c}c\rangle =-\frac{1}{12}\langle \frac{\alpha
_{s}G^{2}}{\pi }\rangle +\frac{1}{m_{c}^{2}}\langle \frac{\alpha _{s}G^{3}}{%
\pi }\rangle \left( -\frac{1}{48}+\frac{13}{720}\right) +\cdots .
\label{eq:Conden}
\end{equation}%
In our analysis we use only the first term in Eq.\ (\ref{eq:Conden})
extracted in Ref.\ \cite{Shifman:1978bx}, because next ones are suppressed
by additional powers of $m_{c}^{-1}$ and are small.

The form factor $g_{3}(Q^{2})$ is evaluated in the domain $Q^{2}=2-20\
\mathrm{GeV}^{2}$: At $Q^{2}=-m_{D^{0}}^{2}$ it equals to $g_{3}$. For $%
(M_{1}^{2},s_{0})$ we have used Eq.\ (\ref{eq:Wind1}), and $%
(M_{2}^{2},s_{0}^{\prime })$ have changed in the regions
\begin{equation}
M_{2}^{2}\in \lbrack 5.5,6.5]~\mathrm{GeV}^{2},\ s_{0}^{\prime }\in \lbrack
34,35]~\mathrm{GeV}^{2}.
\end{equation}%
Results for $g_{3}(Q^{2})$ are depicted in Fig.\ \ref{fig:Fit}. The fit
function $\mathcal{Z}_{3}(Q^{2})$ is fixed by the parameters $\mathcal{Z}%
_{3}^{0}=0.068~\mathrm{GeV}^{-1}$, $z_{3}^{1}=2.46$, and $z_{3}^{2}=3.95.$
The coupling $g_{3}$ is found at $q^{2}=-Q^{2}=m_{D^{0}}^{2}$ and amounts to
\begin{equation}
g_{3}\equiv \mathcal{Z}_{3}(-m_{D^{0}}^{2})=(6.3\pm 1.2)\times 10^{-2}~%
\mathrm{GeV}^{-1}.
\end{equation}
For the width of the process $\mathfrak{M\rightarrow }B^{\ast +}D^{0}$
our analysis predicts
\begin{equation}
\Gamma \left[ \mathfrak{M\rightarrow }B^{\ast +}D^{0}\right] =(9.7\pm 3.0)~%
\mathrm{MeV}.
\end{equation}%
Note that ambiguities above are due to ones in $g_{3}$ and in the masses of
the molecule $\mathfrak{M}$, and mesons $B^{\ast +}$ and $D^{0}$.

The width of the next decay $\mathfrak{M\rightarrow }B^{\ast 0}D^{+}$ is
estimated using the formula
\begin{equation}
\Gamma \left[ \mathfrak{M\rightarrow }B^{\ast 0}D^{+}\right] \approx \Gamma %
\left[ \mathfrak{M\rightarrow }B^{\ast +}D^{0}\right] .
\end{equation}


\subsection{$\mathfrak{M\rightarrow }$ $B^{+}D^{\ast 0}$ and $B^{0}D^{\ast
+} $}


The pair of the decay modes $\mathfrak{M\rightarrow }$ $B^{+}D^{\ast 0}$ and
$\mathfrak{M\rightarrow }$ $B^{0}D^{\ast +}$ is explored in this subsection
using the same techniques. Below, we give parameters of the process $%
\mathfrak{M\rightarrow }$ $B^{+}D^{\ast 0}$ bearing in mind that the second
mode approximately shares those of the first channel.

We consider the correlation function
\begin{equation}
\Pi _{\mu \nu }^{4}(p,p^{\prime })=i^{2}\int d^{4}xd^{4}ye^{ip^{\prime
}y}e^{-ipx}\langle 0|\mathcal{T}\{I^{B^{+}}(y)I_{\mu }^{D^{\ast 0}}(0)I_{\nu
}^{\dagger }(x)\}|0\rangle .  \label{eq:CF2}
\end{equation}%
Here, $I^{B^{+}}(x)$ and $I_{\mu }^{D^{\ast 0}}(x)$ are currents which
correspond to mesons $B^{+}$ and $D^{\ast 0}$, respectively%
\begin{equation}
I^{B^{+}}(x)=\overline{b}_{i}(x)i\gamma _{5}u_{i}(x),\ I_{\mu }^{D^{\ast
0}}(x)=\overline{u}_{j}(x)\gamma _{\mu }c_{j}(x).
\end{equation}%
The correlation function Eq.\ (\ref{eq:CF2}) is necessary to determine the
SR for the form factor $g_{4}(q^{2})$. At the mass shell of the meson $%
D^{\ast 0}$ this form factor allows one to get the strong coupling $g_{4}$
at the vertex $\mathfrak{M}$ $B^{+}D^{\ast 0}$.

We evaluate the function $\Pi _{\mu \nu }^{4\mathrm{Phys}}(p,p^{\prime })$
using the matrix elements
\begin{equation}
\langle 0|I^{B^{+}}|B^{-}(p^{\prime })\rangle =\frac{f_{B}m_{B}^{2}}{m_{b}}%
,\ \ \langle 0|I_{\mu }^{D^{\ast 0}}|D^{\ast 0}(q,\varepsilon )\rangle
=f_{D^{\ast 0}}m_{D^{\ast 0}}\varepsilon _{\mu },  \label{eq:ME4}
\end{equation}%
where $m_{B}=(5279.41\pm 0.07)~\mathrm{MeV}$, $\ f_{B}=206~\mathrm{MeV}$ and
$m_{D^{\ast 0}}=(2006.85\pm 0.05)~\mathrm{MeV}$, $f_{D^{\ast 0}}=(252.2\pm
22.66)~\mathrm{MeV}$ are parameters of the particles under discussion. By $%
\varepsilon _{\mu }$ we denote the polarization vector of the $D^{\ast 0}$
meson. The vertex element $\langle B^{+}(p^{\prime })D^{\ast
0}(q,\varepsilon )|\mathfrak{M}(p,\epsilon )\rangle $ is introduced in the
following form%
\begin{equation}
\langle B^{+}(p^{\prime })D^{\ast 0}(q,\varepsilon )|\mathfrak{M}(p,\epsilon
)\rangle =g_{4}(q^{2})\left[ (p\cdot p^{\prime })(\epsilon \cdot \varepsilon
^{\ast })-(p^{\prime }\cdot \epsilon )(p\cdot \varepsilon ^{\ast })\right] .
\end{equation}

The QCD side of SR is given by the expression
\begin{equation}
\Pi _{\mu \nu }^{4\mathrm{OPE}}(p,p^{\prime })=\frac{1}{3}\int
d^{4}xd^{4}ye^{ip^{\prime }y}e^{-ipx}\langle \overline{c}c\rangle \mathrm{Tr}%
\left[ \gamma _{5}S_{u}^{ij}(y)\gamma _{\mu }S_{c}^{ja}(-x)\gamma _{\nu
}\gamma _{5}S_{b}^{ai}(x-y)\right] .
\end{equation}%
Numerical computations of $g_{4}(Q^{2})$ is performed at $Q^{2}=2-20\
\mathrm{GeV}^{2}$ using the regions Eq.\ (\ref{eq:Wind1}) for $%
(M_{1}^{2},s_{0})$ and
\begin{equation}
M_{2}^{2}\in \lbrack 5.5,6.5]~\mathrm{GeV}^{2},\ s_{0}^{\prime }\in \lbrack
33.5,34.5]~\mathrm{GeV}^{2},
\end{equation}%
for the Borel and continuum subtraction parameters $(M_{2}^{2},s_{0}^{\prime
})$. The extrapolating function $\mathcal{Z}_{4}(Q^{2})$ that has been
employed to compute the coupling $g_{4}(Q^{2})$ is fixed by the following
constants $\mathcal{Z}_{4}^{0}=0.060~\mathrm{GeV}%
^{-1},z_{4}^{1}=3.57,z_{4}^{2}=-2.50$ and shown also in Fig.\ \ref{fig:Fit}.
This function leads to the predictions
\begin{equation}
g_{4}\equiv \mathcal{Z}_{4}(-m_{D^{\ast 0}}^{2})=(5.1\pm 0.9)\times 10^{-2}\
\mathrm{GeV}^{-1},
\end{equation}%
and
\begin{equation}
\Gamma \left[ \mathfrak{M\rightarrow }B^{+}D^{\ast 0}\right] =(6.6\pm 1.7)~%
\mathrm{MeV}.  \label{eq:DW4}
\end{equation}%
The width of the channel $\mathfrak{M\rightarrow }$ $B^{0}D^{\ast +}$ is
also equal to Eq.\ (\ref{eq:DW4}).


\section{Decays to strange mesons}

\label{sec:Widths3}


Decays to strange mesons $\mathfrak{M\rightarrow }B_{s}^{\ast 0}D_{s}^{+}$
and $B_{s}^{0}D_{s}^{\ast +}$ are last two channels of $\mathfrak{M}$ which
have been studied in this article. To consider the decay $\mathfrak{%
M\rightarrow }B_{s}^{\ast 0}D_{s}^{+}$ one should analyze the correlation
function
\begin{equation}
\widehat{\Pi }_{\mu \nu }^{1}(p,p^{\prime })=i^{2}\int
d^{4}xd^{4}ye^{ip^{\prime }y}e^{-ipx}\langle 0|\mathcal{T}\{I_{\mu
}^{B_{s}^{\ast }}(y)I^{D_{s}}(0)I_{\nu }^{\dagger }(x)\}|0\rangle ,
\label{eq:CF3B}
\end{equation}%
which permits one to extract SR for the form factor $G_{1}(Q^{2})$. In Eq.\ (%
\ref{eq:CF3B})$\ I_{\mu }^{B_{s}^{\ast }}(x)$ and $I^{D_{s}}(x)$ are
interpolating currents for the mesons $B_{s}^{\ast 0}$ and $D_{s}^{+}$%
\begin{equation}
I_{\mu }^{B_{s}^{\ast }}(x)=\overline{b}_{i}(x)\gamma _{\mu
}s_{i}(x),~I^{D_{s}}(x)=\overline{s}_{j}(x)i\gamma _{5}c_{j}(x).
\end{equation}%
To determine the physical component $\widehat{\Pi }_{\mu \nu }^{1\mathrm{Phys%
}}(p,p^{\prime })$ of SR, we apply the matrix elements%
\begin{equation}
\langle 0|I_{\mu }^{B^{\ast }}|B_{s}^{\ast 0}(p^{\prime },\varepsilon
)\rangle =f_{B_{s}^{\ast }}m_{B_{s}^{\ast }}\varepsilon _{\mu },\ \ \
\langle 0|I^{D_{s}}|D_{s}^{+}(q)\rangle =\frac{f_{D_{s}}m_{D_{s}}^{2}}{%
m_{c}+m_{s}}.  \label{eq:ME5}
\end{equation}%
In Eq.\ (\ref{eq:ME5}) $m_{B_{s}^{\ast }}=(5415.4\pm 1.4)~\mathrm{MeV}$, $%
f_{B_{s}^{\ast }}=221~\mathrm{MeV}$ and $m_{D_{s}}=(1968.35\pm 0.07)~\mathrm{%
MeV}$, $\ f_{D_{s}}=(249.9\pm 0.5)~\mathrm{MeV}$ are the parameters of the
final-state mesons, and $m_{\mathrm{s}}=(93.5\pm 0.8)~\mathrm{MeV}$ is the
mass of $s$-quark. The vertex $\langle B_{s}^{\ast 0}(p^{\prime
},\varepsilon )D_{s}^{-}(q)|\mathfrak{M}(p,\epsilon )\rangle $ has the
standard form.

\begin{figure}[h]
\includegraphics[width=8.5cm]{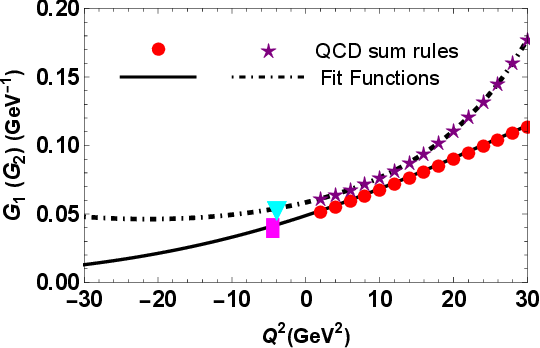}
\caption{SR data for $G_1(Q^{2})$ and $G_{2}(Q^2)$, and extrapolating
functions $\widehat{\mathcal{Z}}_1(Q^{2})$ (solid curve), and $\widehat{%
\mathcal{Z}}_2(Q^{2})$ (dot-dashed curve). Values of these functions at the
points $Q^{2}=-m_{\overline{D}_{s}^{\ast}}^{2}$ and $Q^{2}=-m_{D_{s}}^{2}$
are shown by the rectangle and triangle, respectively.}
\label{fig:Fit1}
\end{figure}

Computation of the correlator $\widehat{\Pi }_{\mu \nu }^{1}(p,p^{\prime })$
in terms of the heavy quark propagators yields
\begin{equation}
\widehat{\Pi }_{\mu \nu }^{1\mathrm{OPE}}(p,p^{\prime })=\frac{1}{3}\int
d^{4}xd^{4}ye^{ip^{\prime }y}e^{-ipx}\langle \overline{c}c\rangle \mathrm{Tr}%
\left[ \gamma _{\mu }S_{s}^{ij}(y)\gamma _{5}S_{c}^{ja}(-x)\gamma _{\nu
}\gamma _{5}S_{b}^{ai}(x-y)\right] .
\end{equation}%
The SR for the form factor $G_{1}(Q^{2})$ is found by employing amplitudes
in $\widehat{\Pi }_{\mu \nu }^{1\mathrm{Phys}}(p,p^{\prime })$ and $\widehat{%
\Pi }_{\mu \nu }^{1\mathrm{OPE}}(p,p^{\prime })$ corresponding to terms $%
\sim g_{\mu \nu }$.

Calculations are carried out using parameters%
\begin{equation}
M_{2}^{2}\in \lbrack 6,7]~\mathrm{GeV}^{2},\ s_{0}^{\prime }\in \lbrack
35,36]~\mathrm{GeV}^{2},
\end{equation}%
for $B_{s}^{\ast 0}$ channel. The constants in the fit function $\widehat{%
\mathcal{Z}}_{1}(Q^{2})$ are $\widehat{\mathcal{Z}}_{1}^{0}=0.0585~\mathrm{%
GeV}^{-1}$, $\widehat{z}_{1}^{1}=2.068$, and $\widehat{z}_{1}^{2}=4.569$.
Results of numerical computations are plotted in Fig.\ \ref{fig:Fit1}, in
which one can see the SR data for $Q^{2}=2-30\ \mathrm{GeV}^{2}$, as well as
the fit function $\widehat{\mathcal{Z}}_{1}(Q^{2})$.

Then the strong coupling $G_{1}$ is equal to
\begin{equation}
G_{1}=\widehat{\mathcal{Z}}_{1}(-m_{D_{s}})=(5.4\pm 1.0)\times 10^{-2}\
\mathrm{GeV}^{-1}.
\end{equation}%
The partial width of the decay $\mathfrak{M\rightarrow }B_{s}^{\ast
0}D_{s}^{+}$ amounts to
\begin{equation}
\Gamma \left[ \mathfrak{M\rightarrow }B_{s}^{\ast 0}D_{s}^{+}\right]
=(6.8\pm 1.8)~\mathrm{MeV}.
\end{equation}

The channel $\mathfrak{M\rightarrow }B_{s}^{0}D_{s}^{\ast +}$ is the next
mode with the strange mesons in the final-state. It is also explored in
accordance with the scheme applied till now. Thus, the correlator necessary
for our analysis is
\begin{equation}
\widehat{\Pi }_{\mu \nu }^{2}(p,p^{\prime })=i^{2}\int
d^{4}xd^{4}ye^{ip^{\prime }y}e^{-ipx}\langle 0|\mathcal{T}\{I_{\mu
}^{B_{s}}(y)I^{D_{s}^{\ast }}(0)I_{\nu }^{\dagger }(x)\}|0\rangle ,
\end{equation}%
where
\begin{equation}
I^{B_{s}}(x)=\overline{b}_{i}(x)i\gamma _{5}s_{i}(x),~I_{\mu }^{D_{s}^{\ast
}}(x)=\overline{s}_{j}(x)\gamma _{\mu }c_{j}(x),
\end{equation}%
are the interpolating currents for mesons $B_{s}^{0}$ and $D_{s}^{\ast +}$,
respectively. The physical side of the sum rule for the form factor $%
G_{2}(q^{2})$ \ is determined by the expression%
\begin{eqnarray}
\widehat{\Pi }_{\mu \nu }^{2\mathrm{Phys}}(p,p^{\prime })&=&\frac{%
G_{2}(q^{2})\Lambda f_{B_{s}}m_{B_{s}}^{2}f_{D_{s}^{\ast }}m_{D_{s}^{\ast }}%
}{(m_{b}+m_{s})\left( p^{2}-m^{2}\right) (p^{\prime 2}-m_{B_{s}}^{2})}\frac{1%
}{(q^{2}-m_{D_{s}^{\ast }}^{2})}\left[ \frac{m^{2}+m_{B_{s}}^{2}-q^{2}}{2}%
g_{\mu \nu }-\frac{m^{2}}{m_{D_{s}^{\ast }}^{2}}p_{\mu }^{\prime }p_{\nu
}^{\prime }\right.  \notag \\
&&\left. +\frac{m^{2}-m_{D_{s}^{\ast }}^{2}}{m_{D_{s}^{\ast }}^{2}}p_{\mu
}^{\prime }p_{\nu }-\frac{m^{2}+m_{B_{s}}^{2}-q^{2}}{2m_{D_{s}^{\ast }}^{2}}%
(p_{\mu }p_{\nu }+p_{\mu }p_{\nu }^{\prime })\right] +\cdots .
\end{eqnarray}%
This correlation function has been derived using the matrix elements
\begin{equation}
\langle 0|I^{B_{s}}|B_{s}^{\ast 0}(p^{\prime })\rangle =\frac{%
f_{B_{s}}m_{B_{s}}^{2}}{m_{b}+m_{s}},\ \ \langle 0|I_{\mu }^{D_{s}^{\ast
}}|D_{s}^{\ast +}(q,\varepsilon )\rangle =f_{D_{s}^{\ast }}m_{D_{s}^{\ast
}}\varepsilon _{\mu },  \label{eq:ME6}
\end{equation}%
and
\begin{equation}
\langle B_{s}^{0}(p^{\prime })D_{s}^{\ast +}(q,\varepsilon )|\mathfrak{M}%
(p,\epsilon )\rangle =G_{2}(q^{2})\left[ (p\cdot p^{\prime })(\epsilon \cdot
\varepsilon ^{\ast })-(p^{\prime }\cdot \epsilon )(p\cdot \varepsilon ^{\ast
})\right] .
\end{equation}%
In Eq.\ (\ref{eq:ME6}) $m_{B_{s}}=(5366.93\pm 0.10)~\mathrm{MeV}$, $%
m_{D_{s}^{\ast }}=(2112.2\pm 0.4)~\mathrm{MeV}$, and $f_{B_{s}}=234~\mathrm{%
MeV}$, $f_{D_{s}^{\ast }}=(268.8\pm 6.5)~\mathrm{MeV}$ are the masses and
decay constants of the final-state particles.

The QCD side $\widehat{\Pi }_{\mu \nu }^{2\mathrm{OPE}}(p,p^{\prime })$ is
given by the formula
\begin{equation}
\widehat{\Pi }_{\mu \nu }^{2\mathrm{OPE}}(p,p^{\prime })=\frac{1}{3}\int
d^{4}xd^{4}ye^{ip^{\prime }y}e^{-ipx}\langle \overline{c}c\rangle \mathrm{Tr}%
\left[ \gamma _{5}S_{s}^{ij}(y)\gamma _{\mu }S_{c}^{ja}(-x)\gamma _{\nu
}\gamma _{5}S_{b}^{ai}(x-y)\right] .
\end{equation}%
Numerical calculations are carried out using the parameters
\begin{equation}
M_{2}^{2}\in \lbrack 5.5,6.5]~\mathrm{GeV}^{2},\ s_{0}^{\prime }\in \lbrack
34,35]~\mathrm{GeV}^{2},
\end{equation}%
in the $B_{s}^{0}$ channel. The parameters of the fit function $\widehat{%
\mathcal{Z}}_{2}(Q^{2})$ are $\widehat{\mathcal{Z}}_{2}^{0}=0.0488~\mathrm{%
GeV}^{-1}$, $\widehat{z}_{2}^{1}=3.495$, and $\widehat{z}_{2}^{2}=-2.438$.
The prediction for $G_{2}$ reads
\begin{equation}
G_{2}=\widehat{\mathcal{Z}}_{2}(-m_{D_{s}^{\ast }})=(4.1\pm 0.8)\times
10^{-2}\ \mathrm{GeV}^{-1}.
\end{equation}%
Then, the partial width of the process $\mathfrak{M\rightarrow }%
B_{s}^{0}D_{s}^{\ast +}$ is

\begin{equation}
\Gamma \left[ \mathfrak{M\rightarrow }B_{s}^{0}D_{s}^{\ast +}\right]
=(3.9\pm 1.1)~\mathrm{MeV}.
\end{equation}

Having utilized results obtained in this work, we estimate the full width of
the molecule $\mathfrak{M}$
\begin{equation}
\Gamma \left[ \mathfrak{M}\right] =(121\pm 17)~\mathrm{MeV},
\end{equation}%
which characterizes it as a broad exotic state. Note that subdominant modes
with the total width $43~\mathrm{MeV}$ form approximately $36\%$ of the
parameter $\Gamma \left[ \mathfrak{M}\right] $.


\section{Discussion and concluding remarks}

\label{sec:Conc}


Parameters of the structure(s) $J/\psi B_{c}^{+}$ [and $\eta _{c}B_{c}^{\ast
+}$] shed light on its properties allowing one to draw some quantitative and
qualitative conclusions about exotic heavy molecular states. The molecules $%
J/\psi B_{c}^{+}$ and $\eta _{c}B_{c}^{\ast +}$ are relatively new members
of the hadron spectroscopy: There are only few works devoted to theoretical
analysis of these systems. Because they have not been yet observed
experimentally, the masses and widths of these structures evaluated in the
present work are important for placing them in due positions inside of this
family.

The mass $m=(9740\pm 70)~\mathrm{MeV}$ has been extracted using the SR
method with a rather high accuracy which means that the mass of the hadronic
molecule is inside of the region $m\in \lbrack 9660,9810]~\mathrm{MeV}$.
This result permits us to conclude that $J/\psi B_{c}^{+}$ and $\eta
_{c}B_{c}^{\ast +}$ lie above corresponding two-meson thresholds and easily
breaks down to constituent mesons, because even in its lower value the mass $%
m=9660~\mathrm{MeV}$ is considerably above the $9372~\mathrm{MeV}$
threshold. In other words, the molecules $J/\psi B_{c}^{+}$ and $\eta
_{c}B_{c}^{\ast +}$ are resonant structures and do not form bound states.
Nevertheless, we treat $J/\psi B_{c}^{+}$ and $\eta _{c}B_{c}^{\ast +}$ as
highly unstable hadronic molecules because they are formed by two heavy
mesons and their interpolating currents are constructed in accordance with
these structures.

It is instructive to compare this situation with the axial-vector
molecule(s) $\mathcal{M}_{\mathrm{AV}}=\Upsilon B_{c}^{-}$ [and $\eta
_{b}B_{c}^{\ast -}$] built of $bb\overline{b}\overline{c}$ quarks. It was
demonstrated that the mass of $\mathcal{M}_{\mathrm{AV}}$ resides in the
region $[15710,15890]~\mathrm{MeV}$, and in the lower border of this
interval $15710~\mathrm{MeV}$ it is stable against decays to mesons $%
\Upsilon B_{c}^{-}$ and $\eta _{b}B_{c}^{\ast -}$. The reason is that this
mass is smaller than corresponding kinematical thresholds $15735~\mathrm{MeV}
$ and $15737~\mathrm{MeV}$ \cite{Agaev:2025wyf}. In this sense the mesons $%
\Upsilon $ and $B_{c}^{-}$, as well as $\eta _{b}$ and $B_{c}^{\ast -}$ may
form bound-state axial-vector molecules. But even in this case they
transform to ordinary mesons due to $\overline{b}b$ annihilations in these
structures.

In other words, the axial-vector state $\Upsilon B_{c}^{-}$ is more stable
against strong decays (in lower mass limit) than the molecule $J/\psi
B_{c}^{+}$. A conclusion about stable nature of some of four-quark mesons
containing $bb$ quarks is not new for the tetraquark spectroscopy. Thus, it
was found that axial-vector diquark-antidiquark state $T_{bb}^{-}=bb%
\overline{u}\overline{d}$ lies below the $B^{-}\overline{B}^{\ast 0}$ and $%
B^{-}\overline{B}^{0}\gamma $ thresholds and is stable against the strong
and radiative transformations \cite%
{Karliner:2017qjm,Eichten:2017ffp,Agaev:2018khe}. Hence, it decays to
conventional mesons via weak processes considered in Ref.\ \cite%
{Agaev:2018khe}. The charmed counterpart of $T_{bb}^{-}$, i.e., a
diquark-antidiquark state $T_{cc}^{+}=cc\overline{u}\overline{d}$ is above a
$D^{\ast }D$ threshold and is not a stable particle \cite{Navarra:2007yw}.
It is remarkable that $T_{cc}^{+}$ was observed by the LHCb collaboration in
$D^{0}D^{0}\pi ^{+}$ mass distribution as a narrow peak with the width $%
\Gamma =(48\pm 2^{+0}_{-14})~\mathrm{keV}$ \cite%
{Aaij:2021vvq,LHCb:2021auc}. The mass of $T_{cc}^{+}$ is very close to the $%
D^{0}D^{\ast +}$ threshold $3875.1~\mathrm{MeV}$, but is smaller than this
limit. This discovery confirmed experimentally that the axial-vector state $%
T_{cc}^{+}$ is strong-interaction unstable structure. The resonance $%
T_{cc}^{+}$ were investigated in the context of diquark and hadronic
molecule models in numerous publications (for details see, Refs.\ \cite%
{Agaev:2021vur,Agaev:2022ast,Ling:2021bir,Feijoo:2021ppq} and references therein).

In general, the two-meson continuum below $m$ may affect predictions of sum
rule computations, because the interpolating current $I_{\mu }(x)$ interacts
with both the molecule $\mathfrak{M}$ and this continuum. Contributions
related to these interactions can be included into analysis by re-scaling
the current coupling $\Lambda $ and keeping $m$ fixed \cite{Agaev:2022ast}.
But their numerical effects are small being around $2-3\%$ of uncertainties
in $\Lambda $ while SR method generates approximately $10\%$ ambiguities.
Therefore, the prediction for $m$ is stable against two-meson contamination.

The full width of $J/\psi B_{c}^{+}$ has been found by considering its
different decay channels. We have computed the partial widths of the
dominant decay modes $\mathfrak{M\rightarrow }J/\psi B_{c}^{+}$ and $%
\mathfrak{M\rightarrow }\eta _{c}B_{c}^{\ast +}$. The subdominant processes,
which imply transformation of $\mathfrak{M}$ to $B^{\ast +}D^{0}$, $B^{\ast
0}D^{+}$, $B^{+}D^{\ast 0}$, $B^{0}D^{\ast +}$, $B_{s}^{\ast 0}D_{s}^{+}$
and $B_{s}^{0}D_{s}^{\ast +}$ mesons, have been explored as well. The
dominant decay channels constitute $64\%$ of $\Gamma \left[ \mathfrak{M}%
\right] $, whereas subdominant modes form remaining $36\%$. Contributions of
subdominant processes may be further refined by taking into account other
possible channels. Our analysis demonstrates that the axial-vector molecule $%
J/\psi B_{c}^{+}$ with the full width $(121\pm 17)~\mathrm{MeV}$ is the
broad resonance above the relevant continuum.

Information on the partial widths of different decay channels as well as
full width of the molecule $\mathfrak{M}$ are important for experimental
analysis of these particles. A broad peak in the mass distributions of
final-state $J/\psi B_{c}^{+}$ or $\eta _{c}B_{c}^{\ast +}$ mesons in some
processes would indicate on observation of all-heavy axial-vector exotic
meson with quark content $cc\overline{c}\overline{b}$. But a
diquark-antidiquark state $cc\overline{c}\overline{b}$ also decays through
these channels. Therefore, a conclusion about internal structure of such a
peak can only be made by comparing experimental results with theoretical
predictions for widths of the molecule and diquark-antidiquark states.
Similarly, some enhancements may be seen in the mass distributions, for
instance, $B^{\ast +}D^{0}$ and $B^{\ast 0}D^{+}$ mesons. But to interpret
them as product of subdominant processes, one first needs to compute
parameters of hadronic molecules $B^{\ast +}D^{0}$ and $B^{\ast 0}D^{+}$
which may dissociate to these mesons as well. Because exotic hadronic
molecules with asymmetric contents are new objects of studies these
problems, in particular decays of such states, are waiting for their
analyses.

Hadronic molecules built of only heavy quarks were on focus of studies in
Ref.\ \cite{Liu:2024pio}. The authors used the extended local gauge
formalism to explore such molecules. They did not find axial-vector bound
states in $cc\overline{c}\overline{b}$ sector, which agree with our
conclusions. But they did not give information on a mass(es) of possible
resonant structures. In the lack of such calculations, it is interesting to
compare our results with predictions made for relevant diquark-antidiquark
states.

The diquark-antidiquark exotic mesons with different spin-parities and
nonsymmetrical quark contents were studied in Refs.\ \cite%
{Liu:2019zuc,Gordillo:2020sgc,Deng:2020iqw,Mutuk:2022nkw,Zhang:2022qtp,An:2022qpt,Galkin:2023wox}%
. In these articles the authors used different methods and approaches such
as nonrelativistic constituent quark, multiquark color flux-tube, and
chromomagnetic models, the Bethe-Salpeter equations and effective potential
model  to investigate fully heavy diquark-antidiquark
structures. Consider, for instance last two publications. The constituent
quark model was used to evaluate parameters of the fully heavy tetraquarks
in Ref.\ \cite{An:2022qpt}, whereas the relativistic quark model employed in
the second paper \cite{Galkin:2023wox}. The mass of the axial-vector
tetraquark $cc\overline{c}\overline{b}$ in these articles were found equal
to $9706~\mathrm{MeV}$ and $9611~\mathrm{MeV}$, respectively. First, one
sees a $100~\mathrm{MeV}$ discrepancy in predictions of different methods.
Nevertheless, we fix the mass gaps around of $34~\mathrm{MeV}$ and $130~%
\mathrm{MeV}$ between diquark-antidiquark and molecular states of the same
content, which are natural for particles with different internal
organizations: Colored diquarks and antidiquarks form more tightly bound
compounds than ordinary colorless mesons.

The molecules $J/\psi B_{c}^{+}$ and $\eta _{c}B_{c}^{\ast +}$ are only
building units of the physical states with asymmetric quark content. Real
particles may be formed owing to mixing of the molecules $\mathfrak{M}$ and $%
\widetilde{\mathfrak{M}}$. Transitions between molecules $J/\psi B_{c}^{+}$
and $\eta _{c}B_{c}^{\ast +}$ as well as their mixing with axial-vector
tetraquark $cc\overline{c}\overline{b}$ are among possible scenarios. But
related problems require additional detailed studies which are beyond of the
current article.

In the present work we have calculated the mass and width of the resonant
state(s) $J/\psi B_{c}^{+}$ [and $\eta _{c}B_{c}^{\ast +}$]. These results
are important to understand the inner structures and properties of the
axial-vector molecules $J/\psi B_{c}^{+}$ and $\eta _{c}B_{c}^{\ast +}$. The
current analysis has clarified only small piece of the whole picture called
"fully heavy exotic mesons". But a lot of efforts should be made to see it
in all details: Comparison of our predictions and alternative ones fills out
blanks in our knowledge about features of these states.

\end{document}